\documentclass[reprint,onecolumn,footinbib,aip,cha,floatfix,longbibliography]{revtex4-1}
\pdfoutput=1

\usepackage{graphicx}
\usepackage[utf8]{inputenc}
\usepackage[english]{babel}

\usepackage{amsmath}
\usepackage{amsfonts}
\usepackage{amssymb}
\usepackage{bm} 

\usepackage{tabularx}
\usepackage{booktabs} 
\usepackage{array} 

\newcommand{\RR}{\mathbb{R}}
\newcommand{\cling}{\mathbf{C}}

\usepackage[colorlinks, breaklinks,hyperfootnotes = false,citecolor = red]{hyperref} 

\begin{document}
\title{Structured networks and coarse-grained descriptions: a dynamical perspective}
\author{Michael T. Schaub}
\email{mschaub@mit.edu}
\affiliation{Institute for Data, Systems and Society, Massachusetts Institute of Technology}
\affiliation{Department of Engineering Science, University of Oxford}
\author{Jean-Charles Delvenne}
\email{jean-charles.delvenne@uclouvain.be}
\affiliation{ICTEAM and CORE, Universit\'e catholique de Louvain}
\author{Renaud Lambiotte}
\email{renaud.lambiotte@maths.ox.ac.uk}
\affiliation{Mathematical Institute, University of Oxford}
\affiliation{naXys, University of Namur}
\author{Mauricio Barahona}
\email{m.barahona@imperial.ac.uk}
\affiliation{Department of Mathematics, Imperial College London}
\affiliation{EPSRC Centre for Mathematics of Precision Healthcare, Imperial College London}

\begin{abstract}\footnotesize
This chapter discusses the interplay between structure and dynamics in complex networks.
Given a particular network with an endowed dynamics, our goal is to find partitions aligned with the dynamical process acting on top of the network.
We thus aim to gain a reduced description of the system that takes into account both its structure and dynamics. 

In the first part, we introduce the general mathematical setup for the types of dynamics we consider throughout the chapter.
We provide two guiding examples, namely consensus dynamics and diffusion processes (random walks), motivating their connection to social network analysis, and provide a brief discussion on the general dynamical framework and its possible extensions.

In the second part, we focus on the influence of graph structure on the dynamics taking place on the network, focussing on three concepts that allow us to gain insight into this notion.
First, we describe how time scale separation can appear in the dynamics on a network as a consequence of graph structure. 
Second, we discuss how the presence of particular symmetries in the network give rise to invariant dynamical subspaces that can be precisely described by graph partitions.
Third, we show how this dynamical viewpoint can be extended to study dynamics on networks with signed edges, which allow us to discuss connections to concepts in social network analysis, such as structural balance.

In the third part, we discuss how to use dynamical processes unfolding on the network to detect meaningful network substructures.
We then show how such dynamical measures can be related to seemingly different algorithm for community detection and coarse-graining proposed in the literature. 
We conclude with a brief summary and highlight interesting open future directions.
\end{abstract}

\maketitle

\section[Introduction]{Introduction}
The language of networks and graphs has become a ubiquitous tool to formalise and analyse systems and relational data across scientific disciplines, from biology to physics, from computer science to sociology~\cite{Newman2010}.
Accordingly, scholars from a variety of areas have investigated such networks from different angles, developing diverse computational and mathematical toolboxes in order to analyse and ascribe meaning to the different patterns found in specific networks of interest.
Modular structures are one of the most commonly studied features of networks in this context~\cite{Schaeffer2007,Fortunato2010,Shai2017,Porter2009,Fortunato2016}.
Yet, as highlighted by the lack of a common terminology (modules, partitions, blocks, communities, and clusters are but a few terms commonly found to denote various notions of modular structure in the literature), \emph{why} scholars are interested in modular structures and \emph{how} these structures are construed can be broadly different.
Hence the perspective adopted when studying the modular structure in networks must depend on the context and specific application in mind~\cite{Shai2017,Schaub2017}.
In the following, we focus on one particular motivation: namely, the rich interplay between network structure and a dynamics acting on top of the network as a means of identifying modules in the network or describing the effect that modules can have on the dynamical behaviour of a system.

\subsubsection*{Why a dynamical perspective?}
One of the main motivations for identifying modular structures in networks is that they provide a simplified, coarse-grained description of the system structure.
Think of a social network, in which we might be able to decompose the system into (overlapping) groups of people such as circles of friends. We may then represent the system in terms of the interactions between these different groups, thereby reducing the complexity of the description.
The hope is not only to arrive at a more compact structural description but also that the obtained modules can be interpreted as `building blocks' with a functional meaning.

For instance, consider the well-known Karate Club network studied by Zachary~\cite{Zachary1977}, representing the social interactions between members of a Karate club that eventually split into two factions after a dispute.
An interesting feature of this network is the fact that the split of the club is commensurate with the graph structure: if we apply graph partitioning methods to this network, the partition into two groups found is commonly well aligned with the split that occurred in reality. While the example of the Karate Club is by no means to be taken as a general indication of the relationship between structure and function, or between network structure and any other type of external data~\cite{Peel2017}, it highlights the ultimate rationale for the detection of modules is often to gain insight into the system behaviour. For instance, we might be interested in how rumours spread in a social network, or opinions are formed.
To understand such processes, we need to take into account the system structure but we also need an understanding of the \emph{dynamics} that acts on top of this structure, since the system behaviour is the result of the interplay between the \emph{structure} of a network \emph{and} the \emph{dynamics} acting on top of it.
We thus aim to gain a reduced description of the system that takes into account both its structure and dynamics.

\subsubsection*{Dynamics \textrm{on} networks or dynamics \textrm{of} networks?}
We should make a distinction here between the dynamics of the network structure itself, which we call \emph{structural dynamics} in the sequel, and dynamical processes that happen on top of a \emph{fixed} network structure.

On the one hand, a social network can be subject to a structural dynamics over time as people become acquainted or start to dislike each other so that links and nodes appear, disappear or change weight
(e.g., if we see who follows whom on twitter, who declares to be friends on Facebook, etc).
The study of how these structures vary over time can be of central importance, e.g., for the spread of pathogens that can spread faster or slower depending on contact patterns. See~\citet{Holme2012} for an overview and further references on these topics.

On the other hand, data may often be naturally interpreted as a dynamics evolving and supported on a latent, unobserved fixed network.
For instance, communication patterns between different people (e.g., on an online social network, an email or a mobile phone call network) may be thought as a type of point process that activates latent links at particular times~\cite{Zhao2015}.
The sequence of activation patterns may not be completely random at each step, but have a certain type of path dependence or memory (e.g., travellers traversing a network of flight connections from one to another city~\cite{Rosvall2014}).
Hence, while the information recorded is temporal, the underlying network itself may be interpreted as a quasi-static object on which a path-dependent dynamics occurs.

There are of course other systems in which the dynamics on the network and the structural dynamics of the network influence each other leading to an evolution of the network structure that reflects the prevalent dynamic patterns on it.
For instance, neuronal networks are known to have high plasticity and adjust their weight structure (links) based on the activity of their nodes (neurons), a feature that is commonly associated with learning.

Whether one should focus on structural dynamics, dynamics on top of a network, or both is therefore dependent on what the network representation aims to capture.
In reality, all of these viewpoints are ultimately abstractions and thus attempts to capture different aspects of real world systems which hopefully provide additional insight into their behaviour.

\subsubsection*{Network dynamics --- the scope of this chapter}
Our focus here will be on dynamical processes acting on top of networks. 
We thereby assume that the underlying (latent) network structure is known and approximately constant over the time scales of the observed dynamics. 
Hence, we largely omit the issue of structural dynamics, even though this may not be justified in certain applications.
In the sequel, we will show that this approach is fruitful in many contexts, yielding insights that go beyond purely structural network analysis.
While clearly important, the joint treatment of structural dynamics in conjunction with dynamics on networks has received less attention in the literature and requires a more elaborate mathematical machinery that goes beyond the scope of this chapter.
Furthermore, we do not consider here the question of how and why the networks have arisen in the first place.
(A reader interested in these questions may refer to some of the other chapters in this book.)
We will therefore assume that the observed network is well defined, i.e., we treat it as an empirical reality with low uncertainty. The dynamical perspective adopted here is especially useful in such cases: the network is specified, but the emergent behaviour (our object of interest) might be hard to grasp due to the complexity of the system.

More explicitly, think again of the Karate club example.  From a statistical perspective, 
one might want to answer the question of why the structure of the network is as observed.
We may adopt a generative model (e.g., a stochastic blockmodel) and assume that the observed network is a random realisation from this model.
We could then attempt to find a classification of the nodes such that the observed link probabilities between blocks of nodes 
reflect the observed block structure, hence explaining parsimoniously the main features in the data~\cite{Snijders2011}.
Using this perspective, we assume that if we could repeat the `experiment' that created the network multiple times, the realisation of the network would be different each time, and we want our model to correspond to the simplest generative process consistent with those observations.
In many circumstances this is a hypothetical question, however, as we only have access to a single observed network, and thus need to assume that our class of models (e.g., stochastic blockmodels) provides a suitable approximate depiction of all important features of the network. 

Here we ask a complementary question: given the particular network we observe and an endowed dynamics taken place on it, are there partitions aligned with this process? 
For the Karate club this could give an indication of whether the split of the club was facilitated by how its particular network structure influenced the opinion formation process in this social network.
Irrespective of the network's genesis, these types of questions are of interest in many areas and underpin our perspective in this chapter.

\subsubsection*{Outline of this chapter}
We divide this chapter into three parts.
In the first part, we introduce the general mathematical setup for the types of dynamics we consider throughout the chapter.
We provide two guiding examples, namely consensus dynamics and difussion processes (random walks), motivating their connection to social network analysis, and provide a brief discussion on the general dynamical framework and its possible extensions.

In the second part, we focus on the influence of graph structure on the dynamics taking place on the network, focussing on three concepts that allow us to gain insight into this notion.
First, we describe how time scale separation can appear in the dynamics on a network as a consequence of graph structure. 
Second, we discuss how the presence of particular symmetries in the network give rise to invariant dynamical subspaces that can be precisely described by graph partitions.
Third, we show how this dynamical viewpoint can be extended to study dynamics on networks with signed edges, which allow us to discuss connections to concepts in social network analysis, such as structural balance.

In the third part, we discuss how to use dynamical processes unfolding on the network to detect meaningful network substructures. We then show how different such measures can be related to seemingly different methods for community detection and coarse-graining proposed in the literature. 
We conclude with a brief summary and highlight interesting open future directions. 

Our account is geared towards conveying intuition rather than covering technical details. We provide pointers to additional literature with detailed results throughout the text.

\subsubsection*{Notation}
For simplicity, in the following we consider mainly undirected, connected graphs with $n$ nodes (vertices) and $m$ links (edges). 
Our ideas extend to directed graphs, however, and we provide appropriate references to the literature for the interested reader as we go along.
The topology of a graph is encoded in the weighted adjacency matrix $\mathbf{A}\in \RR^{n\times n}$, where $A_{ij}$ is the weight of the link between node $i$ and node $j$.
Clearly, for an undirected graph $\mathbf{A}= \mathbf A^\top$.
Typically, most graphs are unsigned (i.e., $A_{ij} \geq 0, \, \forall i,j$).
The weighted out-degrees (or strengths) of the nodes are given by the vector $\mathbf{d} = \mathbf A \mathbf{1}$, where $\mathbf{1}$ is the $n \times 1$ vector of ones.
For a given vector $\mathbf{v}$, we will sometimes define the associated diagonal matrix $\textrm{diag}(\mathbf{v})$ with elements $v_i$ on the diagonal and zero elsewhere. 
For instance, we define the diagonal matrix of degrees $\mathbf D = \text{diag}(\mathbf{d})$ and denote the total weight of the edges by $w =  \mathbf{1}^\top \mathbf{D} \mathbf{1}/2 = \mathbf{1}^\top \mathbf{d} / 2$.

The combinatorial graph Laplacian is defined as $\mathbf L=\mathbf D-\mathbf A$, while the normalised graph Laplacian is defined as $\mathbf L_\mathrm{N} =\mathbf D^{-1/2} \mathbf L \mathbf D^{-1/2} = \mathbf I-\mathbf D^{-1/2} \mathbf A \mathbf D^{-1/2} $.  
Both these Laplacians are symmetric positive semi-definite, with a simple zero eigenvalue when the graph is connected~\cite{Chung1997,Godsil2013}.
When describing diffusion processes on graphs, it is also useful to define the (asymmetric) random walk Laplacian $\mathbf L_\mathrm{RW} = \mathbf D^{-1} \mathbf L$, which is isospectral with the normalised Laplacian for undirected graphs.

We will also consider \textit{signed graphs}, where the weights $A_{ij}$ can be positive or negative. 
In the case of signed graphs, we define the vector of absolute degrees 
$\mathbf{d}_\mathrm{S} = |\mathbf A| \mathbf{1}$, where the absolute value is taken element-wise,
with the corresponding absolute degree matrix $\mathbf D_\mathrm{S} = \text{diag}(\mathbf{d}_\mathrm{S})$.
For signed networks, we will define the signed Laplacian $\mathbf L_\mathrm{S} = \mathbf D_\mathrm{S}-\mathbf A$, which is also positive semi-definite. The signed Laplacian reduces to the combinatorial Laplacian in the case of an unsigned graph. 

A (hard) partition $\mathcal P$ of a graph of $n$ nodes into $k$ cells $\{\mathcal C_i\}^k_{i=1}$ can be encoded by an indicator matrix $\cling \in {\{0,1\}}^{n \times k}$, with entries $C_{ij} =1$ if node $i$ is part of cell $\mathcal C_j$ and $C_{ij}=0$ otherwise.  
Hence the columns of $\cling$ are the indicator vectors $\mathbf{c}^{(i)}$ of the cells: 
\begin{align}
\label{eq:H_cols}
\cling& := [\mathbf{c}^{(1)},\ldots,\mathbf{c}^{(k)}].
\end{align}

\section{Part I -- Dynamics on and of networks}\label{sec:partI}

\subsection{General setup}
In its most general form, we are interested in dynamical systems of the form:
\begin{subequations}
\label{eq:gen_setup}
\begin{align}
    \mathbf{\dot{x}}(t)&= \mathcal A \left(\mathbf{x}(t), t \right) \, \mathbf{x}(t) + \mathcal B(t) \, \mathbf{u}(t) & \quad  \mathcal{A}\in \mathbb{R}^{\ell \times \ell }, \quad \mathcal{B}\in\mathbb{R}^{\ell\times p}&\\
    \mathbf{y}(t) &= \mathcal D(t) \, \mathbf{x}(t) & \quad \mathcal{D} \in \mathbb{R}^{c \times \ell}&,
\end{align} 
\end{subequations}
where $\mathbf{x} \in \mathbb{R}^{\ell} , \mathbf{y}  \in \mathbb{R}^{c}, \mathbf{u}  \in \mathbb{R}^{p}$ are the state, the observed state, and the input vectors of the system, respectively. 
Discrete time versions are also of interest~\cite{Delvenne2010,Delvenne2013}, but we will stick to the continuous time version henceforth.

In the context of networked systems, the system of ODEs~\eqref{eq:gen_setup} arises by endowing each node with one or more state variables, whose union corresponds to the state vector $\mathbf x$.
In general, the matrix $\mathcal A$ is linked to the `network': a time-varying, state-dependent coupling between the state variables of the agents (nodes). A set of exogenous inputs, described by the vector $\mathbf u$ acts on the state variables through the input matrix $\mathcal B$. In such a system, we may not be able to observe and measure all the system states. This is captured by the fact that the output $\mathbf y$ is a linear transformation of $\mathbf x$.
This framework can naturally account for weighted, signed or other types of interactions. Furthermore, the fact that each node can be endowed with several state variables allows for the modelling of higher order dynamics (e.g., higher order Markov processes)~\cite{Rosvall2014,Salnikov2016,Hoffmann2012,Delvenne2015}.
Note that this form also allows for the inclusion of exogenous inputs, a factor usually neglected in standard network analyses, although it has recently gained prominence for the problem of controlling networks~\cite{Liu2011}.

The system~\eqref{eq:gen_setup} formally describes the full coupled dynamics \textit{of} and \textit{on} a network, since the `network' (encapsulated in the matrix $\mathcal A \left(\mathbf{x}(t), t \right)$) is both state and time-dependent. However, such systems are difficult to analyze in general. When the coupling $\mathcal{A}(t)$ is only time-dependent, the system describes the dynamics on a time-varying network .
Such linear time-varying models have a long history in systems and control theory, and there is a rich literature pertaining to their analysis~\cite{Kailath1980,Brockett2015} at the expense of more advanced mathematical machinery.
Although a growing literature in \textit{dynamical social network analysis} melding such concepts from control and dynamical systems with social network analysis (e.g., for opinion formation~\cite{Proskurnikov2017}) has recently emerged,
such models have been comparably less studied within the scope of network theory. 

\subsubsection*{Dynamics on fixed networks}
To simplify our exposition, we will here assume that the (latent) coupling remains constant over time, 
i.e., $\mathcal{A,B,D}$ have no explicit time dependence. This is what we have termed \textit{dynamics on a (fixed) network}. It is important to remark that this assumption does not imply that each link is constantly activated over time, but that it is available for a potential interaction~\cite{Rosvall2014,Salnikov2016,Hoffmann2012,Delvenne2015}.
We will also assume that $\ell=p=c=n$, which implies that there is only one state variable per node:
\begin{subequations}
\label{eq:simple_setup}
\begin{align}
    \mathbf{\dot{x}}&= \mathcal A \, \mathbf{x} + \mathcal B \, \mathbf{u}(t) & \quad  \mathcal{A} \in \mathbb{R}^{n \times n}, \quad \mathcal{B} \in \mathbb{R}^{n \times n}&\\
    \mathbf{y} &= \mathcal D \, \mathbf{x} & \quad \mathcal{D} \in \mathbb{R}^{n \times n}.
\end{align} 
\end{subequations}

In the following, we consider examples of this simpler form, which provide rich insights into problems of interest in practical applications. Specifically, we first consider consensus dynamics and its variants (motivated by opinion formation), followed by diffusion processes and random walk dynamics (motivated by information propagation).

\subsection{Consensus dynamics}
\begin{figure}[tb!]
    \centering
    \includegraphics[width=\textwidth]{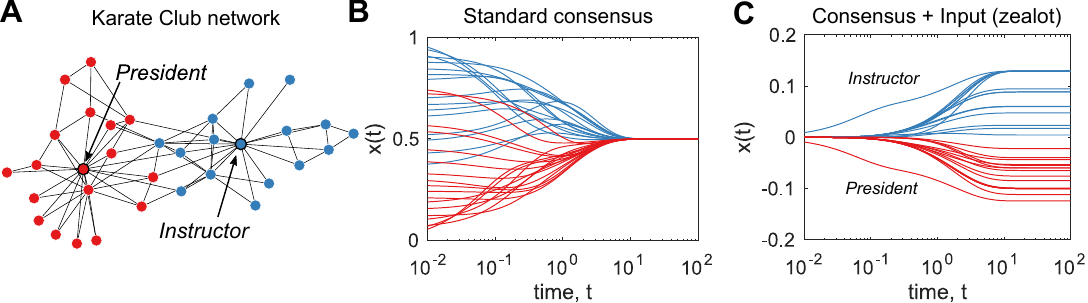}
    \caption{\textbf{Consensus dynamics on the Karate Club network.} \textbf{A}~The Karate Club network originally analysed by Zachary~\cite{Zachary1977} with nodes coloured according to the split that occurred in the real case. \textbf{B}~Consensus dynamics on the Karate club network starting from a random initial condition. As time progresses, the states of the individual nodes become more aligned and eventually reach the consensus value equal to the arithmetic average of the initial condition. Note that above the time scale given by the eigenvalue $1/\lambda_2(\mathbf{L}) \approx 1/0.47$, the agents converge into two groups that reflect the observed split before converging to global consensus  (see Section~\ref{sec:timescale}).
    \textbf{C} If an external input is applied to the system (see text), the opinion dynamics will in general not converge to a single value but lead to a dispersed set of final opinions, which still reflect the split observed in reality. }%
    \label{fig:example_consensus}
\end{figure}

Consensus is one of the most popular and well studied dynamics on networks~\cite{Jadbabaie2003,Olfati-Saber2004,Ren2005,Olfati-Saber2007,Yu2011,Ren2007}, and can be thought of as a linear version of synchronisation~\cite{Barahona2002,Jadbabaie2004}.
The attractiveness of consensus lies in its analytic tractability and simplicity, which nevertheless provides a good first description of some fundamental behaviours.  
For instance, in the socio-economic domain, consensus provides a model for 
opinion formation in a society of individuals, whereas in engineering systems, consensus constitutes a basic building block for efficient distributed computation of global functions in networks of sensors, robots, or other agents~\cite{Jadbabaie2003}.
For a recent survey of consensus processes with a particular focus on opinion formation, we refer the reader to Proskurnikov et al.~\cite{Proskurnikov2017}.

To define the standard consensus dynamics, consider a given connected network of $n$ nodes with adjacency matrix $\mathbf A$.
Let us endow each node with a scalar state variable $x_i \in \mathbb{R}$.
The consensus dynamics on such a network is defined as:
\begin{equation} \label{eq:consensus}
    \mathbf{\dot{x}} = - \mathbf L \mathbf{x}, \qquad \text{(consensus dynamics)}
\end{equation}
where $\mathbf L$ is the graph Laplacian. Clearly, the consensus dynamics amounts to 
$$\dot{x}_i = -D_{ii}x_i+ \sum_j A_{ij}x_j = - \sum_j A_{ij}(x_i-x_j), \, \forall i,$$ 
i.e., each node adjusts its state such that the difference to its neighbours is reduced.
The name of the dynamics derives from the fact that for any given initial state $\mathbf{x_0} = \mathbf{x}(0)$, the differential equation~\eqref{eq:consensus} drives the state to a global `consensus state', where the state variables of all nodes are equal to the arithmetic average of the initial node states: $x_i = x_*  \,\, \forall i$,  where $x_* = \mathbf{1}^\top \mathbf{x_0} / n$  as $t\rightarrow \infty$. 
Relative to our framework~\eqref{eq:simple_setup}, the standard consensus dynamics~\eqref{eq:consensus} corresponds to $\mathcal{A}=- \mathbf{L}, \, \mathcal{D}= \mathbf{I}, \, \mathbf{u}=\mathbf{0}$.

Intuitively, this dynamics may be interpreted as an opinion formation process on a network of agents who, in the absence of further inputs, will eventually agree on the same value of their state (`opinion'), namely, the average opinion of their initial states. 
Figure~\ref{fig:example_consensus}B shows an example of the consensus dynamics on the Karate Club network starting from a random initial condition for the agents and converging asymptotically towards the common, final `opinion'. Yet the network structure plays a role in the form in which this opinion is approached: the opinions of each of the two factions (as recorded by their eventual split in real life) converge earlier towards a `group opinion' with higher cohesion.

While in the absence of external inputs the standard consensus dynamics converges to a fixed point, the framework~\eqref{eq:simple_setup} allows us to explore the influence of inputs over time $\mathbf{u}(t) \neq \mathbf{0}$, e.g., by external agents, media, etc. 
In that case, the asymptotic convergence of the dynamics to an eventual consensus is not guaranteed. For instance, some agents may behave like `zealots', who do not update their opinion as described above, but give more weight to their own opinion~\cite{Mobilia2007,Proskurnikov2016,Acemoglu2013}.
Let us consider the Karate Club network with a constant external input:
\begin{equation} 
    \mathbf{\dot{x}} = - \mathbf L \mathbf{x} + \mathbf{u},
\end{equation}
with $u_\text{president} =-1$ for the president, $u_\text{instructor}=+1$ for the instructor, and all other nodes have no input $u_i=0$. This can be thought of as a simplified model of a zealot-like behaviour of these two agents.
In this case, there is no final consensus reached within the system: the final opinion of each of the agents is dispersed between the extreme positions taken by the instructor and the president (Figure~\ref{fig:example_consensus}C).
Importantly, the final opinions of the agents are well aligned with the split that eventually occurred in the Karate club, in which half of the members joined the instructor to form a new club and the other half stayed with the president.

These results highlight how the graph properties (encapsulated by the graph Laplacian $L$) can shape and constrain the dynamics on the network, and thus influence the observed behaviour of the system. 

\subsubsection*{Discussion: More detailed consensus models} 
The consensus dynamics studied here is chosen for its simplicity. Of course, in real world systems the process of opinion formation is much more complex.
For instance, opinions can be interlinked and part of a belief system~\cite{Proskurnikov2017,Friedkin2016}; update and gossiping processes may be nonlinear or asynchronous~\cite{Jadbabaie2004, Olfati-Saber2007,Yu2011}; and noisy external inputs may influence the process~\cite{Young2010}. All these factors lead to a much more complex dynamics.
In particular, opinions may not converge to a single value, or might stabilise to different values in different parts of the network. See~\citet{Proskurnikov2017} and references therein for a discussion on the so-called social cleavage problem.

\subsection{Diffusion processes and random walks}

Random walks are another important dynamical process which can naturally evolve on graphs.  
Random walks are often taken as a (simple) proxy for diffusive processes and, like consensus processes, these type of models have found applications in various domains, including information diffusion in social networks.
Other applications of such processes include searching on networks, web browsing, dimensionality reduction via diffusion maps, respondent driven sampling, and, indeed, community detection~\cite{Aldous2002,Masuda2016}.
Perhaps the most popular example is the celebrated PageRank algorithm, which is used for the ranking of webpages, and can be seen as an application of random walks on directed networks. 

The dynamics of a continuous-time unbiased random walk on a network with combinatorial Laplacian $\mathbf{L}$ is governed by the Kolmogorov forward equation:
\begin{equation} \label{eq:cont_random_walk}
     \dot{\mathbf{p}}^\top = -\mathbf{p}^\top \mathbf{L}_\mathrm{RW}
    \qquad \text{(random walk)}
\end{equation}
where $\mathbf{L}_\mathrm{RW} = \mathbf D^{-1} \mathbf L$ is the random walk Laplacian.
This equation describes the diffusion of a random particle on the network; specifically, the time evolution of the probability mass function $\mathbf p(t)$ of an $n$-dimensional random vector 
$\mathbf{x}(t)$ with components $x_i(t) = 1$ if the particle is at node $i$ at time $t$ and zero otherwise. 
For a connected, undirected network, this dynamics will converge to the stationary distribution $\bm{\pi} = \mathbf{d}/2w$. 
For the general case of a directed graph, see~\cite{Delvenne2013,Lambiotte2009, Lambiotte2014}. 

\begin{figure}[tb!]
    \centering
    \includegraphics[width=\textwidth]{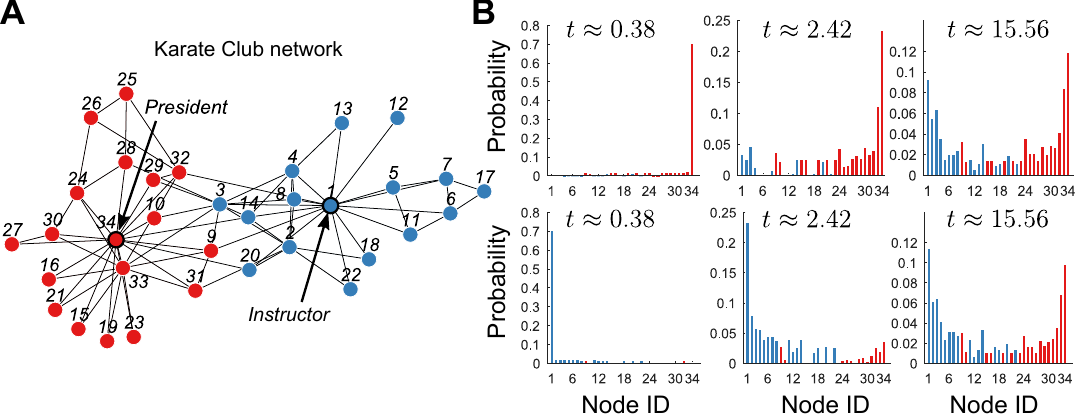}
    \caption{\textbf{Illustration of the evolution of a random walk dynamics on the Karate Club network.} \textbf{A} The Karate Club network with labelled nodes. The factions of the ultimate split observed in reality are indicated by color (grayscale). 
        \textbf{B} The evolution of the probability distribution of the random walk over time exemplified at 3 time snapshots from two different initial conditions: the random walker starts at time $t=0$ at the `president' node 34 (upper three panels), or  at the `instructor' node 1 (lower three panels).
    As time evolves, the probability of the walker to be found on the other nodes becomes more spread out on the graph and eventually reaches the stationary distribution $\bm{\pi}=\mathbf{d}/2w$.  Note that for short times, the probability is spread mostly within the corresponding `factions' (i.e., president for the top panels; instructor for the bottom panels). Beyond the slowest time scale in this dynamics given by $1/\lambda(\mathbf{L}_\text{RW}) \approx 1/0.13$, the random walk becomes well mixed and the information spreads across factions (see Section~\ref{sec:timescale} for more details).}%
    \label{fig:example_random_walk}
\end{figure}

An illustration of the time evolution of such a random walk on the Karate club network is displayed in Figure~\ref{fig:example_random_walk}. 
As time progresses, the process converges towards the distribution $\bm{\pi}$, which is proportional to the degree vector.
It is known that the degree is also a simple centrality measure for the nodes of a graph~\cite{Lambiotte2014}, and this observed behaviour highlights how the centrality of the instructor and the president (the highest degree nodes) is also of dynamical importance.
Figure~\ref{fig:example_random_walk}B also illustrates the notion of information propagation: two random walks started one from the instructor or alternatively from the president spread initially mostly within their natural groups before eventually spreading across the whole network.

\subsubsection*{Discussion: Random walks and consensus as dual processes} It is worth remarking that the random walk dynamics~\eqref{eq:cont_random_walk} may be seen as the dual of a non-standard consensus dynamics of the form $\mathbf{\dot{x}} = -\mathbf D^{-1}\mathbf L \mathbf{x}$, which (unlike the standard consensus~\eqref{eq:consensus}) converges to a different final value: a \textit{degree-weighted} average of the initial node-states.
Conversely, a combinatorial Laplacian dynamics of the form $\mathbf{\dot{p}}^\top = -\mathbf{p}^\top \mathbf L$ can also be seen as describing a different random walk, which is the dual of the standard consensus~\eqref{eq:consensus}.

\section{Part II -- The influence of graph structure on network dynamics}
\label{sec:partII}

\subsection{Time scale separation in partitioned networks}
\label{sec:timescale}

\subsubsection*{Standard time scale separation}

The classic concept of time scale separation in a dynamical system follows from the simple system:
\begin{subequations}
\label{eq:time_sep}
\begin{align}
    \frac{dx}{dt} = f(x,y), \label{eq:time_sep1} \\
    \epsilon^{-1} \frac{dy}{dt} = g(x,y),  \label{eq:time_sep2}
\end{align}    
\end{subequations}
where $f,g$ are bounded functions and $\epsilon$ is a small constant relative to those bounds. 
By a simple rescaling we may assume that $f, g$ are of order 1 and $\epsilon \ll 1$.
In this system, $x(t)$ changes much more rapidly than $y(t)$ since $dy/dt = \epsilon g(x,y)$, which is small by construction.
An alternative, equivalent statement follows from defining a `slow' time variable $\tau := \epsilon \, t$, hence ~\eqref{eq:time_sep2} can be rewritten as $dy/d\tau = g(x,y)$.
From this rewriting, it follows that there is a \emph{separation of time scales} in the dynamics, where $y$ evolves according to the slow time scale $\tau$, and $x$ evolves according to the faster $t$.

When a time scale separation is present in a system, it can be exploited to simplify its analysis in various ways.
If we are interested in the short term behaviour of the system, we may effectively treat $y$ as a fixed parameter and ignore its time evolution, leading to an effective one-dimensional system description based on the \textit{fast} time scale, for instance describing the (fast) convergence of $x$ to the fixed point $x_*(y)$.
On the other hand, if we are interested in the long term behaviour of the system, then it is $y$ we are most interested in.
Since the dynamics of $x$ is faster than that of $y$, we may assume that $x \simeq x_*(y)$ at all times beyond an initial transient, leading to a one-dimensional system for the evolution of $y(t)$.
Using these simplifications will, of course, lead to errors when comparing the approximation to the 
actual time-evolution. However, the error can be bounded through time scale separation theory.

In summary, when there are two well separated time scales in the system ($t$ and $\tau = \epsilon t$) the dynamics `almost decouples' into two different regimes: for the fast behaviour, we may simply concentrate on $x$, whereas for the slow, long-term behaviour we may focus on $y$ and forget about the detailed dynamics of $x$.  
When several \emph{distinct} time scales are present, we can similarly approximate the dynamics over particular time scales by reduced dynamics that can be obtained by finding quasi-invariant subspaces in the original system. 
These concepts emerge naturally in the study of networked dynamics, as we discuss below.

\subsubsection*{Time scales in consensus dynamics on networks}
Time scale separation can appear naturally in the context of dynamics on networks.
For simplicity, we will describe the results here in the context of consensus dynamics, though translating these ideas to diffusion processes is straightforward.  

Given an initial condition $\mathbf{x}_0$, linear systems theory tells us that the solution to the consensus dynamics~\eqref{eq:consensus} is given by 
$$\mathbf{x}(t) = \exp(-\mathbf Lt) \, \mathbf{x}_0,$$ 
where $\exp(\cdot)$ denotes the matrix exponential.
Writing the full solution in this manner obscures the time scales present in the evolution of $\mathbf{x}(t)$, and how they get mixed via the network interactions represented by the Laplacian $\mathbf L$.
To reveal the characteristic time scales, we use the spectral decomposition of the Laplacian (if the graph is undirected), or a slightly more careful, yet related, treatment for directed graphs~\cite{Lambiotte2014}.

For simplicity of exposition, let us assume that the graph is undirected (i.e., $\mathbf L = \mathbf L^\top$).
Let us denote the eigenvectors of $\mathbf L$ by $\mathbf{v}_i$ with associated eigenvalues $\lambda_i$
in increasing order $0=\lambda_1 \le \lambda_2 \le \ldots \le \lambda_n$.
The spectral decomposition of the Laplacian is then: 
$$\mathbf L = \sum_{i=1}^n \lambda_i \mathbf{v}_i^{} \mathbf{v}_i^\top.$$  Accordingly, the solution of the consensus dynamics can be written in the spectral basis as 
$$\mathbf{x}(t) =  \sum_{i=1}^n e^{-\lambda_i t} \, \mathbf{v}_i^{}\mathbf{v}_i^\top\mathbf{x}_0 =  \sum_{i=1}^n \left(\mathbf{v}_i^\top\mathbf{x}_0 \right ) e^{-\lambda_i t} \, \mathbf{v}_i. $$

In this rewriting, the time scales of the process become apparent: they are dictated by the eigenvalues of the Laplacian matrix, with each eigenvector (eigenmode) decaying with a characteristic time scale $\tau_i = 1/\lambda_i$.
Hence, if there are large differences (gaps) between eigenvalues, the system will have time scale separation.
For instance, if the $k$ smallest eigenvalues $\{\lambda_1=0,\ldots,\lambda_k\}$ are well separated from the remaining eigenvalues such that $\lambda_k \ll \lambda_{k+1}$, the eigenmodes associated with 
$\{\lambda_{k+1}, \ldots, \lambda_n\}$ become negligible for $t  > 1/\lambda_{k+1}$ and it follows that the system can be effectively described by the $k$ smallest eigenmodes for  $t > 1/\lambda_{k+1}$.
More technically, we say that the first  $k$ eigenvectors form a dominant invariant subspace of the dynamics and there exists an associated lower dimensional ($k < n$) description of the dynamics on the network which is valid after the time scale $1/\lambda_{k+1}$. 
A natural question is: how is the time scale separation that emerges from the spectral properties of the Laplacian related to the network structure?  In the case of networks, time scale separation is typically associated with a lower dimensional description of the dynamics which is aligned with \emph{localised substructures} in the graph, as we illustrate through the following example.

\subsubsection*{Example: A modular partitioned network structure induces time scale separation}
\begin{figure}[tb!]
    \centering
    \includegraphics[]{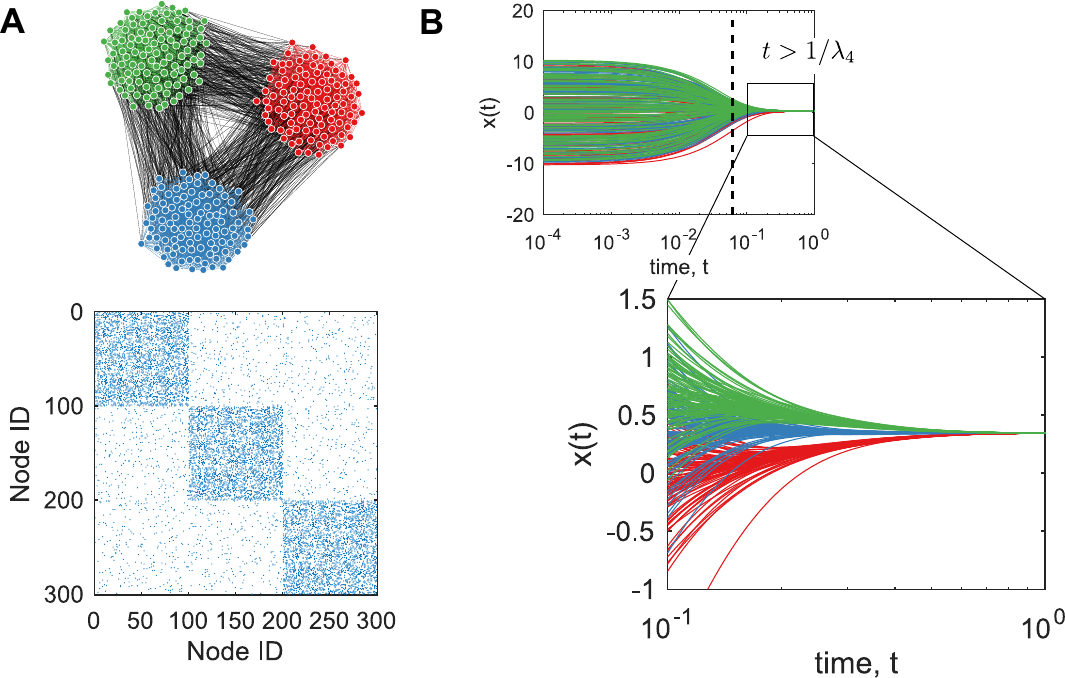}
    \caption{\textbf{Consensus dynamics on a structured network.} \textbf{A} Visualisation of the network and adjacency matrix of an unweighted structured network with 3 groups of the form~\eqref{eq:block_network}. \textbf{B}~The consensus dynamics on this network displays time scale separation: after $t \approx 1/ \lambda_4 = 0.06$, approximate consensus is reached within each group (groups indicated by color/grayscale) followed by global consensus across the whole network. 
A similar effect can be observed in the consensus dynamics and random walk on the Karate club network in Figures~\ref{fig:example_consensus}~ and~\ref{fig:example_random_walk}}.%
    \label{fig:consensus_structured}
\end{figure}

To illustrate the discussion above, let us consider a network composed of $k$ modules with strong in-block coupling and weaker inter-block coupling, as given by the adjacency matrix
\begin{equation}
\label{eq:block_network}
    \mathbf A = 
    \begin{pmatrix}
        \mathbf A_1 &  & & \\
            & \mathbf A_2 & & \\
            & & \ddots & \\
            & & & \mathbf A_k
    \end{pmatrix} + 
    \mathbf A_\text{random} := \mathbf A_{\text{structure}}+ \mathbf A_\text{random}.
\end{equation}
Here $\mathbf A_\text{random}$ can be interpreted as a realisation of an Erd\H{o}s-R\'enyi (ER) random graph with unstructured, sparse connectivity, whereas the $\mathbf{A}_i$ are the adjacency sub-matrices of the individual modules that have higher random connectivity inside. Here, we model these modules also as ER graphs that possess a higher connectivity probability than $\mathbf A_\text{random}$ (Figure ~\ref{fig:consensus_structured}).
How does the structure present in the graph affect the spectrum and eigenvectors of the corresponding Laplacian, $\mathbf L = \mathbf L_\text{structure} + \mathbf L_\text{random}$?

Let us first consider the case where $\mathbf A_\text{random} = 0$, i.e., the graph consists of $k$ disconnected components.
In that case, it is easy to see that we have a repeated eigenvalue $\lambda = 0$ with 
multiplicity $k$ and the associated $k$-dimensional eigenspace can be spanned by the $k$ indicator vectors $\mathbf{c}^{(j)} \in \mathbb{R}^{n}$ defined in Eq.~\eqref{eq:H_cols}, which are localised on the blocks in the graph.  

To gain insight into the case where $\mathbf{A}_\text{random} \neq 0$, we invoke matrix perturbation theory and random matrix theory~\cite{VonLuxburg2007,Stewart2001,Benaych-Georges2011,Handel2017}. For a network of the form~\eqref{eq:block_network}, a form of the Davis-Kahan theorem~\cite{Rohe2011,Bandeira2015,Lei2015,Handel2017} provides bounds on the angular distance between the subspace $\mathcal S$ spanned by the block-vectors $\{\mathbf{c}^{(j)}\}_{i=1}^k$ and the subspace $\mathcal S'$ spanned by the eigenvectors of $\mathbf L$ associated with the $k$ smallest eigenvalues.

Intuitively, Davis-Kahan states that if the random component is small, then $\mathcal S \approx \mathcal S'$. The implication is that the dominant invariant subspaces will be commensurate with the structural decomposition of the network in terms of the block-vectors. Hence the long-term dynamics will directly reflect the structural decomposition of the network. 
In other words, the time scale separation in such a networked system takes an intuitive meaning: quasi-consensus is reached more quickly within each block, while global consensus is only reached on a longer time scale.

These points are illustrated numerically in Figure~\ref{fig:consensus_structured}, where we show how the consensus dynamics evolves on a network of the form~\eqref{eq:block_network} consisting of 3 groups with 100 nodes each. 
As shown in Figure~\ref{fig:consensus_structured}B, the dynamics becomes effectively low dimensional after around $t  \simeq 1/\lambda_4$, beyond which it is well approximated by the 3 dominant eigenmodes aligned with the intrinsic blocks in the network.

\subsubsection*{Discussion: Time scale separation beyond homogeneous block structures}
It is important to emphasise that time scale separation may also be induced by network structures that are not block-homogeneous. Many networks contain natural `non-clique like' substructures (e.g., ring-shaped), which may act effectively as a dynamical substructure over a particular time scale~\cite{Schaub2012,Schaub2015}. The presence of such substructures will too affect the observed consensus dynamics on the network.

Furthermore, our discussion is not limited to the case where the structure of the network is block-diagonal, but can be extended seamlessly to networks consisting of a low-rank structure plus a random `noise' component~\cite{Benaych-Georges2011,Nadakuditi2012,Peixoto2013,Zhang2014}. Such networks encompass stochastic blockmodels~\cite{Rohe2011,Lei2015}, although the spectral properties of a realisation of a stochastic blockmodel may not be concentrated around their expectation when the network is very sparse~\cite{Mossel2013,Decelle2011,Krzakala2013}. 
For more details on stochastic blockmodels, see~\citet{Snijders2011,Anderson1992,Holland1983} and some of the other chapters in this book.

\subsection{Strictly invariant subspaces of the network dynamics and external equitable partitions}

Let us now consider another type of network structure that induces a specific form of exact coarse-graining of the dynamical process acting on the network~\cite{OClery2013,Schaub2016a}: the so-called \emph{external equitable partition} (EEP).

In order to introduce EEPs, we first recall the well known graph-theoretic notion of equitable partition~\cite{Godsil2013}.
An equitable partition splits the graph into a set $\{\mathcal C_i\}$ of non-overlapping groups of nodes called cells, that fulfil the following condition:  \textit{for each node $v\in\mathcal C_i$, the number of connections to nodes in cell $\mathcal C_j$ is only dependent on $i,j$.}  
Stated differently, the nodes inside each cell of an equitable partition have the same out-degree pattern with respect to every cell. 
The EEP is a relaxed version of the equitable partition:  the requirement on equal out-degree need only hold for the number 
of connections between \textit{different} cells $\mathcal C_i, \mathcal C_j$, where $i \neq j$.  
EPs and EEPs are closely related to so-called orbit partitions and to symmetry properties of a graph, and may be detected using tools from computational group theory~\cite{Sorrentino2016,Pecora2014,Sanchez-Garcia2018}.
An example of a graph with an EEP is shown in Figure~\ref{fig:EEP_dynamics}A.  

The presence of an EEP in a network has important consequences for a dynamics taking place on it. 
To see this, we consider the algebraic characterisation~\cite{cardoso2007laplacian,Egerstedt2012,Chan1997} of an EEP of a graph of $n$ nodes into $k$ cells encoded by the $n \times k$ indicator matrix $\cling$~\eqref{eq:H_cols}. 
Associated with the EEP there is a quotient graph---a coarse-grained version of the original graph, such that each cell becomes a node and the weights between these new nodes are the total out-degrees between the cells in the original graph (Fig.~\ref{fig:EEP_dynamics}A).
It can then be verified that 
\begin{equation}
\label{eq:EEP_def}
    \mathbf L\cling = \cling \mathbf{L}^\pi,
\end{equation}
where $\mathbf{L}^\pi$ is the $k \times k$ Laplacian of the quotient graph induced by $\cling$: 
\begin{equation}
\label{eq:Lpi}
    \mathbf{L}^\pi =  (\cling^\top \cling)^{-1}\cling^\top   \mathbf L\cling = \cling^+\mathbf L\cling.
\end{equation}
Here the $k \times n$ matrix $\cling^+$ is the (left) Moore-Penrose pseudoinverse of $\cling$.
Observe that multiplying a vector $\mathbf{x} \in \mathbb R^n$ by $\cling^\top $ from the left sums up the components within each cell, and that $\cling^\top \cling$ is a diagonal matrix with the number of nodes per cell on the diagonal.
Hence, $\cling^+= (\cling^\top \cling)^{-1}\cling^\top$ can be simply interpreted as a \emph{cell averaging operator}~\cite{OClery2013}. 
After straightforward algebraic manipulations it is easy to show that:
\begin{equation}
\label{eq:H+_commutator}
  \cling^+\mathbf L =  \mathbf{L}^\pi \cling^+,
\end{equation}
which summarises the relationship between the cell averaging operator $\cling^+$ and the Laplacians of the original and quotient graphs.
Note that although the Laplacian of the original (undirected) graph $\mathbf L$ is symmetric, the Laplacian of the quotient graph $\mathbf{L}^\pi$ will be asymmetric in general. 

\subsubsection*{Dynamical implications of EEPs}
The definition of the EEP~\eqref{eq:EEP_def} implies an invariance of the partition encoded by $\cling$ with respect to the Laplacian $\mathbf{L}$. Specifically, if we apply the Laplacian to the indicator matrix $\cling$ we obtain a linearly rescaled (by $\mathbf L^\pi$) version of $\cling$.
A similar invariance of the cell averaging operator $\cling^+$ with respect to $\mathbf L$ underpins Eq.~\eqref{eq:H+_commutator}.

Let us expand on some of the consequences of this invariance.
Equation~\eqref{eq:EEP_def} implies that the columns of $\cling$ span an invariant subspace of $\mathbf L$.
As the invariant subspaces of $\mathbf L$ are expressible in terms of its eigenvectors, it follows that there exists a set of eigenvectors of $\mathbf L$ whose components are constant on each cell of the partition.
Furthermore, it can be shown that the eigenvalues associated with the eigenvectors spanning the invariant subspace are shared with $\mathbf{L}^\pi$~\cite{OClery2013}. 
If $\mathbf L$ has degenerate eigenvalues, an eigenbasis can be chosen such that it is localised on the cells of the partition~\cite{OClery2013}.

\begin{figure}
\includegraphics[]{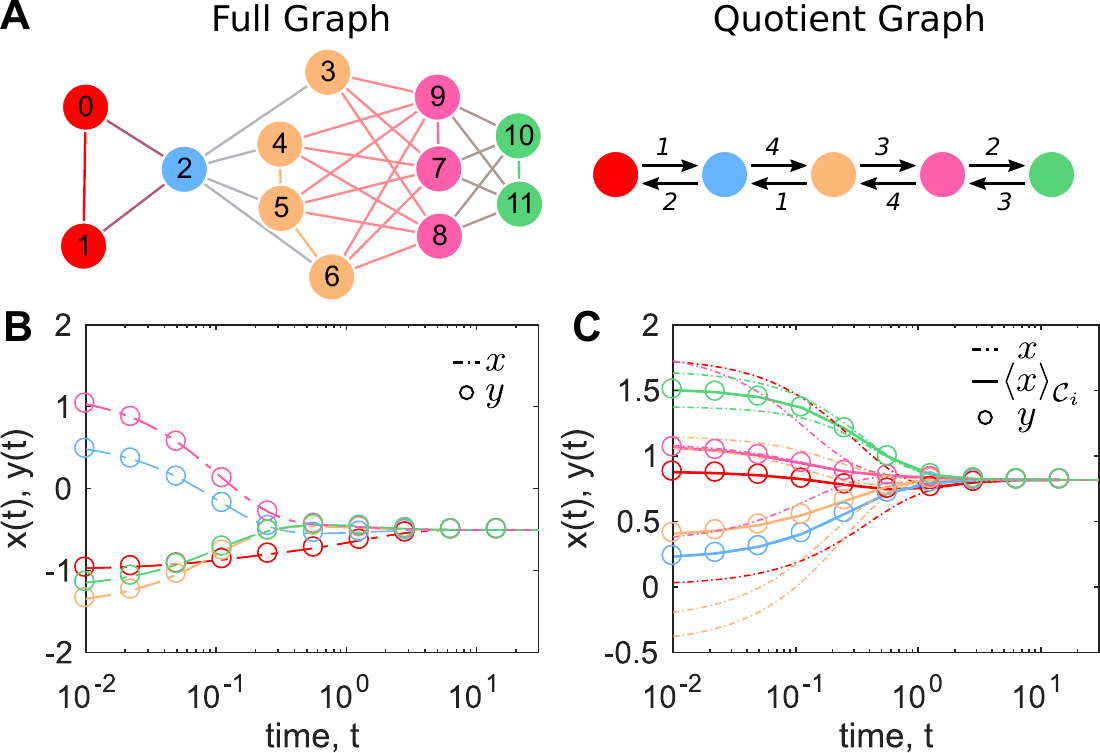}
\caption{\textbf{The external equitable partition and its dynamical implications.} 
   \textbf{A}  A graph with $n=12$ nodes (left) with an external equitable partition into five cells (indicated with colors) and its associated quotient graph according to the EEP (right).
\textbf{B} Invariance of the EEP: The consensus dynamics on the full graph~\eqref{eq:consensus} from an initial condition $\mathbf{x}=\cling\mathbf{y}$ is shown with dash-dotted lines, whereas the associated quotient dynamics~\eqref{eq:invariant_cells} governing $\mathbf{y}$ is shown with circles. If all states within each cell are equal (i.e., cluster-synchronised), the dynamics will always remain cluster-synchronised and are described by the dynamics of the quotient graph for all times.
\textbf{C} Cell-averaging dynamics of the EEP: For consensus dynamics, the quotient graph dynamics (circles) also describes the cell-averaged dynamics (solid lines) of the unsynchronised full graph dynamics (dash-dotted lines), as given by~\eqref{eq:cell_averages}. 
}
\label{fig:EEP_dynamics}
\end{figure}

These algebraic properties of an EEP have implications for the dynamics dictated by $\mathbf L$, as we illustrate now for consensus dynamics~\cite{OClery2013}.
First, an EEP is consistent with \textit{partial consensus} such that the agreement within a cell (if present) is preserved.
Specifically, let the initial state vector be given by $\mathbf{x} = \cling \mathbf{y}$ for some arbitrary $\mathbf{y}$, so that every node within cell $\mathcal{C}_i$ has the same initial value $y_i$ of their `opinion' variable.
Then the nodes inside each cell remain identical \textit{for all times} and the dynamics of the cell variables is governed by the quotient graph:
\begin{align}
\label{eq:invariant_cells}
    \dot{\mathbf{x}} & = \cling\dot{\mathbf{y}} \quad \text{with} \quad     
\dot{\mathbf{y}}  = - \mathbf{L}^\pi \mathbf{y},
\end{align}
which follows directly from Eq.~\eqref{eq:EEP_def}. 
In words, if consensus has been reached within each EEP cell, then the lower dimensional Laplacian matrix $\mathbf L^\pi$ (with dimensionality equal to the number of cells $k$) describes
the full dynamics of the system (Fig.~\ref{fig:EEP_dynamics}B). This dynamical invariance thus provides a simpler model of the system.

A second consequence of the presence of an EEP is that the dynamics of the cell-averaged states 
$\langle  \mathbf{x} \rangle_{\mathcal{C}_i} $
is exactly described by the quotient graph:
\begin{equation}
\label{eq:cell_averages}
  \frac{d  \langle \mathbf{x} \rangle_{\mathcal{C}_i}}{dt} = 
    -\mathbf{L}^\pi \langle \mathbf{x} \rangle_{\mathcal{C}_i} \quad \text{where} \quad {\langle  \mathbf{x} \rangle_{\mathcal{C}_i} := \cling^+  \mathbf{x}}.
\end{equation}
This dynamical coarse-graining follows directly from Eq.~\eqref{eq:H+_commutator}.
Hence the cell-averaged dynamics is also governed by the lower dimensional quotient Laplacian (Fig.~\ref{fig:EEP_dynamics}C), and if we are only interested in such cell averages we can reduce our model significantly.

Finally, a third implication of the EEP structure relates to the system with inputs. It can be shown~\cite{OClery2013} that all the results  for the autonomous consensus dynamics with no inputs can be equivalently rephrased for the system with inputs:  
\begin{equation}\label{eq:consensus_input}
    \dot{\mathbf{x}} =-\mathbf L\mathbf{x} + \mathbf{u}(t),
\end{equation}
when the input ${\mathbf{u}(t) = \cling\mathbf{v}(t)}, {\mathbf{v}(t)\in \mathbb R^c}$ is consistent with the cells of an EEP. 
In that case, the nodes inside each cell remain identical for all times, as in~Eq.\eqref{eq:invariant_cells}. 

\paragraph*{\textbf{Remark:}} 
While we have focused here on the implications of an EEP for linear consensus dynamics, invariant partitions like the EEP play a similar role for other linear and \textit{nonlinear} dynamics (e.g., Kuramoto synchronisation). See Ref.~\cite{Schaub2016a} for an extended discussion including synchronisation dynamics, as well as dynamics on signed networks.

\subsubsection*{Discussion: Differences and relationships between EEPs and time scale separation}
Let us discuss briefly the difference between the presence of an EEP in a network, and time scale separation in the system. 
In our context, both concepts can be related to the notion of (strictly or almost) invariant subspaces in the dynamics. 
However, the link between structure and dynamics that each of them represents can be very different.

The presence of an EEP is related to symmetries in the graph, which translate into the fact that a set of Laplacian eigenvectors have components that are constant on each cell in the graph. 
However, these eigenvectors can be associated to any eigenvalue of the graph, whether fast or slow.
In broad terms, for the EEP the `shape' of the eigenvectors with respect to the cells is important, but the eigenvalues themselves are not relevant.

This notion is therefore different to the time scale separation discussed in Section~\ref{sec:timescale}, where the defining criterion focusses on the eigenvalues---more precisely, on the existence of gaps between eigenvalues that separate them into groups associated with different time scales.
In our particular example of a planted partition model in Figure~\ref{fig:consensus_structured}, the associated eigenvectors were indeed approximately constant on each cell (i.e., on each block of nodes) and would tend to align with the cells as the random part decreases. Hence in this case both the (approximate) EEP structure and the time scale separation are well aligned.
However, this may not always the case.
We may indeed have an EEP in which the set of eigenvectors corresponding to the cells are precisely the slowest eigenmodes, but this is not necessary.
Conversely, the eigenvectors corresponding to the slowest time scales do not have to be exactly constant on every cell, 
or may not be block-structured in general~\cite{Schaub2012,Schaub2015}.  Therefore the notions of EEP and time scale separation are distinct but not mutually exclusive.

\subsection{Structural balance: consensus on signed networks and polarised opinion dynamics}

\label{sec:signed}

\subsubsection*{Signed networks and structural balance}
In social networks, relationships can be friendly or hostile, or reflect either trust or distrust between individuals.  
The positive or negative character of links between agents is a central concept associated with the emergence of conflict and tension in social psychology, and has been studied within the classic literature in social network theory~\cite{Heider1946,Cartwright1956}. More recently, the study of networks with \textit{signed interactions} has gained popularity in the context of online social networks \cite{Leskovec2010} and online cooperation \cite{Szell2010}. More broadly, networks with signed edges are also essential to model biological systems and their dynamics. Examples include genes that promote or repress the expression of other genes in genetic regulatory networks~\cite{Davidson2005}, or neurons that can excite or inhibit the firing of other neurons in neuronal networks, thereby shaping the global dynamics of the system~\cite{Gerstner2014,Schaub2015}.

\begin{figure}
\center
\includegraphics{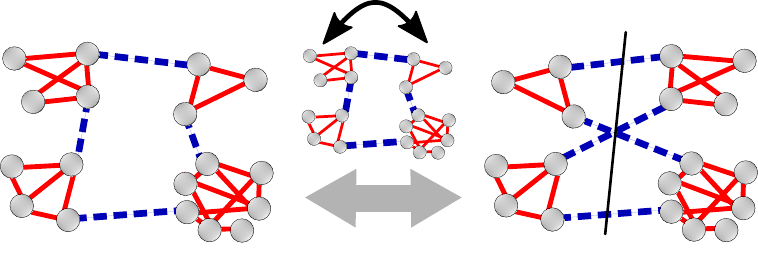}
\caption{\textbf{Example of a structurally balanced graph.}  A  structurally balanced graph (left) with positive edges (solid, red), and negative edges (dashed, blue) can be redrawn in a different way (right) so that it can be split into two groups such that the negative edges connect nodes of different groups and links between nodes in the same group are positive.}
\label{fig:struct_balanced_graph}
\end{figure}

Following Cartwright~\cite{Cartwright1956}, a network is \textit{structurally balanced} if the product of the signs along any closed path in the network is positive. This implies that only consistent social relationships are allowed in triangles of three nodes: either all interactions are positive, or there are exactly two negative links, which may be interpreted in the sense that ``the enemy of my enemy is my friend''~\cite{Heider1946}.
Equivalently, a network is structurally balanced if it can be split into two factions, where each faction contains only positive interactions internally, while the connections between the two factions are purely antagonistic~\cite{Altafini2013}, as
illustrated in Figure \ref{fig:struct_balanced_graph}.
It has been shown that many social networks are empirically close to structural balance~\cite{Facchetti2011}, suggesting that there might be a structural dynamics \textit{driving} social networks towards structural balance~\cite{Marvel2011,Traag2013}.

\subsubsection*{Consensus dynamics on signed networks and polarisation}

There are several mathematical formulations that incorporate positive and negative network interactions. 
Here we consider systems where the interactions are mediated through a signed Laplacian matrix defined as~\cite{Altafini2013,Kunegis2010}:
\begin{equation}
\mathbf L_\mathrm{S} = \mathbf D_\mathrm{S} - \mathbf A,
\end{equation}
where $\mathbf D_\mathrm{S} = \text{diag}(|\mathbf A|\mathbf{1})$ is the diagonal absolute degree matrix, 
and the adjacency matrix $\mathbf A$ can contain both positive and negative weights.
Like the standard Laplacian, the signed Laplacian $\mathbf L_\mathrm{S}$ is positive semidefinite and its spectrum contains a zero eigenvalue when the graph is connected and structurally balanced~\cite{Altafini2013,Kunegis2010}.
To see this, note that the signed Laplacian can be decomposed as:
\begin{equation}
    \mathbf L_\mathrm{S} = \mathbf B^{} \mathbf W_\text{abs} \mathbf B^\top,
\end{equation}
where $\mathbf W_\text{abs} =  \text{diag}(|\mathbf w_e|)$ is the diagonal matrix containing the absolute edge weights and $\mathbf B \in \mathbb R^{n \times m}$ is the node-to-edge incidence matrix:
\begin{equation*}
    B_{ie} = \begin{cases} 1& \text{ if $i$ is the tail of edge $e$}, \\ -\text{sign}(e)& \text{ if $i$ is the head of edge $e$.}\end{cases}
\end{equation*}
When the graph is connected but not structurally balanced,  $\mathbf L_S$ is positive definite.

The following interesting alternative characterisation of a structurally balanced graph was highlighted by Altafini~\cite{Altafini2013,Altafini2013a}: a network is structurally balanced if there exists 
a $\pm 1$ diagonal matrix $\bm{\Sigma}$ such that the matrix
\begin{equation}
    \mathbf L_{\bm \Sigma} = \bm \Sigma \mathbf L_\mathrm{S} \bm \Sigma
\end{equation}
contains only negative elements on the off-diagonal.  
In other words,  $ \mathbf L_{\bm \Sigma}$ is a standard Laplacian matrix for another (associated) graph with only positive weights.
The matrix $\bm \Sigma$ is called switching equivalence, signature similarly, or gauge transformation in the literature~\cite{Altafini2013a}.
Choosing $\bm \Sigma = \mathbf{I}$ makes it clear that a network with only positive weights is itself balanced, which emphasises that $\mathbf L_\mathrm{S}$ is a proper generalisation of the standard graph Laplacian.
Using this characterisation, one can efficiently determine whether a network is structurally balanced~\cite{Facchetti2011} and obtain the corresponding switching equivalence matrix $\bm \Sigma$.

We are now in a position to study the opinion dynamics on a signed network given by:
\begin{equation}\label{eq:signed_consensus}
    \dot{\mathbf{x}} = -\mathbf L_\mathrm{S}\mathbf{x}.
\end{equation}
The above equation is often called `signed consensus' dynamics and we will adopt this name here, but it is 
important to note that this dynamics does not have to lead to a single consensus for all agents~\cite{Altafini2013,Altafini2012,Proskurnikov2016}.
Intuitively, this should be clear: since the agents can be repelled by the opinions of their neighbours, the eventual dynamics need not converge to a unique consensus, even in the absence of external inputs.
This is in contrast to the standard consensus case shown in Figure~\ref{fig:example_consensus}. 

\begin{figure}
\centering
\includegraphics{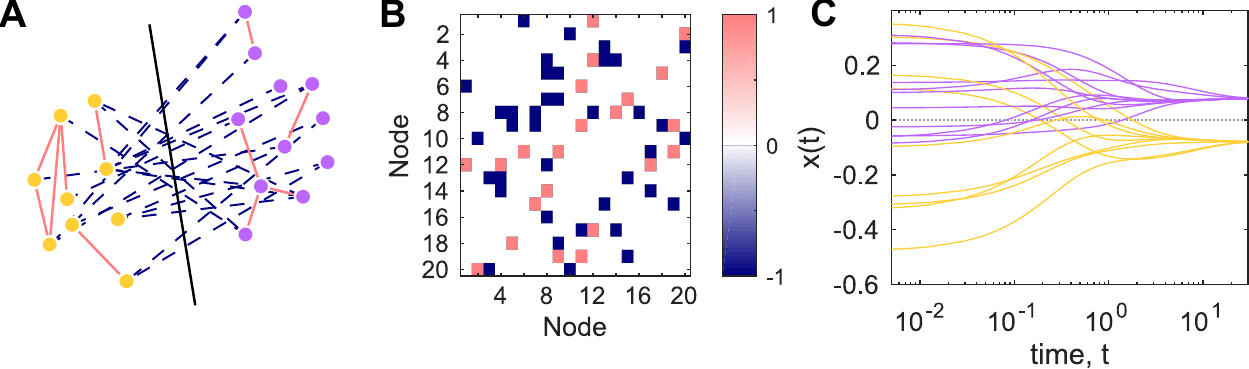}
\caption{\textbf{Polarised opinion dynamics in structurally balanced networks.} \textbf{A} Visualisation of a structurally balanced network with positive (solid) and negative (dashed) links. As indicated by the drawing, the network can be partitioned into two subsets such that all negative links are between nodes in different sets. \textbf{B} The adjacency matrix of the same structurally balanced network. Fom this representation it is no immediately obvious whether or not the network is structurally balanced. \textbf{C} Starting from a random initial condition, the signed consensus dynamics~\eqref{eq:signed_consensus} on this structurally balanced network converges asymptotically to a state of two polarised opinions~\eqref{eq:final_state_signed_consensus}, where the final opinions are exactly the same in magnitude, but opposite in sign for the two groups shown in \textbf{A}.  }%
\label{fig:polarised_dynamics}
\end{figure}

A remarkable feature of structurally balanced networks is that the signed consensus dynamics~\eqref{eq:signed_consensus} converges asymptotically to a \textit{polarised state}, i.e., a state where the final value for all nodes is the same in magnitude, but the nodes are divided into two sets with opposite sign (Figure~\ref{fig:struct_balanced_graph}). 
More precisely, there is an eigenvector of $\mathbf L_\mathrm{S}$ associated with eigenvalue $0$ of the form $\bm \sigma = [\sigma_1,\ldots,\sigma_N]^T$ where all $\sigma_i \in \{-1,+1\}$ such that the dynamics~\eqref{eq:signed_consensus} converges to the final state 
\begin{equation}\label{eq:final_state_signed_consensus}
    \lim_{t\rightarrow \infty} \mathbf x(t) = \dfrac{\bm \sigma^T \mathbf x_0}{n}\bm \sigma.
\end{equation}
The sign pattern of $\bm \sigma$ corresponds precisely to the switching equivalence transformation $\bm \Sigma = \text{diag }(\bm \sigma)$ and $\sigma_i$ is denoted the polarisation of node $i$. 
Note that the overall sign of $\bm \sigma$ is arbitrary, so that the polarisation of a node is only meaningful relative to that of other nodes.
An example of this behaviour is shown in Figure~\ref{fig:polarised_dynamics}.

\subsubsection*{Discussion: Dynamics of networks towards structural balance}  
Our discussions up to now have focussed mostly on the effects of graph structure on the dynamics taking place on a fixed network. Let us now briefly discuss a form of structural dynamics (where the network structure itself changes) in the context of signed networks. 

Albeit commonly explained from a static perspective (i.e., parity of the cyclic paths in the network), 
structural balance is essentially a dynamical theory---it posits that networks tend to \emph{evolve} towards a state of structural balance.
An important question is therefore how such a structural dynamics of network evolution could look like.
Antal et. al~\cite{Antal2005} provided one of the first explorations of this issue in the discrete time setting, followed by 
the continuous time version by Kulakowski et al.~\cite{Kulakowski2005}, which was
analysed in detail by Marvel and collaborators~\cite{Marvel2011}.
Here we focus on the following variation of this model proposed by Traag et al.~\cite{Traag2013}:
\begin{equation}\label{eq:traag_model}
    \dot{\mathbf{X}} = \mathbf{X}\mathbf{X}^\top \qquad \text{with} \qquad \mathbf{X}(0) = \mathbf{X}_0,
\end{equation}
where $\mathbf{X}\in \mathbb R^{n\times n}$ denotes the matrix of opinions that agents have of each other. 

It can be shown~\cite{Traag2013} that this model converges to a structurally balanced network for almost all initial conditions, 
i.e., we reach either a split into two factions of opposing opinions, or a state in which all nodes have positive opinion about each other.
More precisely, the normalised solution of Eq.~\eqref{eq:traag_model} converges for some time $t^*$ to:
\begin{equation*}
    \lim_{t \rightarrow t^*} \dfrac{\mathbf{X}(t)}{\|\mathbf{X}\|_F} = \mathbf{v}\mathbf{v}^T,
\end{equation*}
where $\|\cdot\|_F$ denotes the Frobenius norm and $\mathbf{v}$ is a real-valued vector whose sign pattern indicates the two factions.  
However, an important shortcoming of this model is the fact that while the \textit{sign pattern} converges to a balanced network, the \textit{magnitude} of the opinions diverge after a finite time $t^*$, unless certain technical conditions are fulfilled~\cite{Traag2013}.
A second shortcoming is the assumption of homogeneity in the network an in the agents: each agent has an opinion about every other agent in the network (`all-to-all' connectivity) and all agents follow the same update rule. These assumptions are unrealistic in many real world scenarios. However, extended models that relax these simplifications are far more difficult to analyse rigorously, thus providing an interesting object for further study.

\section{Part III -- Using dynamical processes to reveal network structure}\label{sec:partIII}

As we have seen in the previous sections, graph structure can have a notable impact on a dynamics acting on a network.
What about the converse?  Can we choose a dynamics, let it evolve on a graph, and learn a modular or coarse-grained description of the network based on some of its properties (time scale  separation, quasi-invariance, etc)?

In the following, we provide a high-level, algorithmic point of view of how this may be achieved.
Interestingly, many community detection methods and graph partitioning heuristics in the literature 
can be seen as particular cases of this general abstract viewpoint~\cite{Delvenne2010,Delvenne2013}.
Although not every method for community detection is best interpreted in terms of dynamical quantities~\cite{Schaub2017},
the dynamics-based perspective presented here can serve as a general framework to establish differences and similarities between the plethora of existing methods. Our perspective reveals some surprising relationships between measures and methods that have been proposed in seemingly different ares in the literature.

\begin{figure}[tb!]
    \centering
    \includegraphics[width=\textwidth]{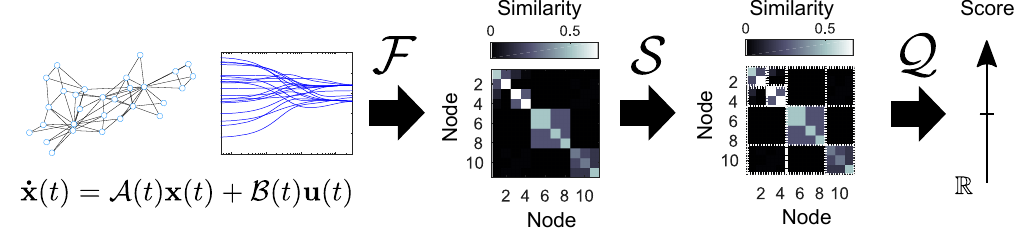}
    \caption{\textbf{Schematic of the algorithmic framework for partitioning of networks based on dynamics.} A dynamics $\mathbf{x}(t)$ operating on a network with inputs $\mathbf{u}(t)$ is filtered through the inference operator $\mathcal{F}$ to provide a similarity matrix between the nodes, from which a partitioning (or coarse-graining) of the nodes is obtained by a selection and summary operator $\mathcal{S}$. The quality of this partition is summarised by a numerical metric through the quality operator $\mathcal{Q}$. This abstraction of dynamics-based community detection covers many existing methods in the literature, and opens up a wide array of possibilities suited to different purposes.}%
    \label{fig:schematic_framework}
\end{figure}

\subsection{A generic algorithmic framework for dynamics-based network partitioning and coarse graining}

Given a complex network, one important task is to find relevant communities (i.e., groups of nodes that can act as a coarse-grained description of the particular network) in an unsupervised manner. There is a very large array of such methods.
However, as we will make more precise through an example below, one can interpret many of the community detection algorithms through the following generic `algorithmic recipe', which includes four ingredients:
\begin{enumerate}
    \item A \textit{dynamical process} $\mathbf x: t \mapsto \mathbf x(t)$ acting on a given network. 
    \item An \textit{inference operator} $\mathcal F$, which is used to characterise some statistical information about the dynamical process. 
    \item A \textit{selection and summary operator} $\mathcal S$, which assigns nodes into groups and clusters (or coarse-grains) the dynamical process.
    \item A \textit{quality function} $\mathcal Q$, which scores the clustering 
\end{enumerate}

While more general scenarios can be used, in the cases we discuss below these ingredients take the following form.
Given a graph, we use a dynamical process $\mathbf x(t)$ to assign to each node a particular time trajectory or a time-evolving probability measure.
We then apply the inference operator $\mathcal F$ on this time-course data to produce a node-to-node time-dependent similarity measure. The similarity measure is then aggregated according to a given grouping of nodes by the selection operator $\mathcal S$.
The proposed clustering is then evaluated by the quality function $\mathcal Q$, which can then be optimised (Figure \ref{fig:schematic_framework}).

\subsubsection*{Example of the framework: Markov Stability for community detection}
Let us exemplify these ideas through the Markov Stability framework, as discussed in Ref.~\cite{Delvenne2010,Delvenne2013,Schaub2012}.
We use this example to demonstrate the applicability of the proposed algorithmic framework, since 
it can be shown that the Markov Stability framework encompasses a variety of community detection methods as particular cases~\cite{Delvenne2010,Delvenne2013,Schaub2012}, including modularity, spectral clustering, and various Potts models.
For a discussion that includes processes beyond diffusion, and further relations to control-theoretic notions we refer to~\citet{Schaub2018}.

Markov Stability is a quality function that scores a given non-overlapping partition $\mathcal P$ 
of a graph into node groups $\mathcal C_i$.  
The Markov Stability~\cite{Delvenne2010,Lambiotte2009} score is parametrically dependent on a time parameter $t$ and is given by:
\begin{equation}\label{eq:stability_prob}
    \text{Markov Stability} (t,\mathcal P) := r(t, \mathcal P)  = \sum_{\mathcal C_i \in \mathcal P} \text{Pr}(\mathcal C_i,0,t) - \text{Pr}(\mathcal C_i,0,\infty)
\end{equation}
where $\text{Pr}(\mathcal C_i,0,t)$ is the probability for a random walker to be in the same cell $\mathcal C_i$ at time zero and at time $t$.  
For an extended discussion of this framework, we refer to further literature~\cite{Delvenne2010,Delvenne2013,Schaub2012, Lambiotte2009, Lambiotte2014}.

We now see how Markov Stability fits within the above four ingredient `recipe':
\begin{enumerate}
\item To an undirected, connected graph with Laplacian $\mathbf{L}$ we associate a diffusion dynamics (random walk) given by: 
\begin{equation} \label{eq:random_walk_2}
    \dot{\mathbf{p}}^\top = -\mathbf{p}^\top \mathbf{L}_\mathrm{RW},
\end{equation}
which describes the evolution of the probability mass function $\mathbf p(t)$ of an 
$n$-dimensional random indicator vector $\mathbf{x}(t)$ with $x_k(t) = 1$ 
if the particle is at node $i$ at time $t$ and zero otherwise. Recall that $ \mathbf{L}_\mathrm{RW} = \mathbf D^{-1} \mathbf L$
is the random walk Laplacian.

\item The inference operator $\mathcal F$ is set to be the \textit{autocovariance function} of this process, which computes the correlation between the random state vector $\mathbf{x}(\tau)$ at stationarity with itself after a time $t$, $\mathbf{x}(\tau + t)$):
\begin{equation}
\label{eq:autocovariance}
 \mathcal F( \mathbf{x}(t)) = \text{cov}\left[ \mathbf{x}(\tau),\mathbf{x}(\tau\!+\!t)\right] = \bm \Pi \mathbf P(t) -\bm{\pi}\bm{\pi}^\top.
\end{equation}
$\bm{\pi} = \mathbf{d}/2w$ denotes the stationary distribution of the diffusion. 

\item For a given partition of the graph with indicator matrix $\mathbf{C}$, we choose a selection and summary operator $\mathcal S$ that aggregates the entries of an $n \times n$ matrix defined on the nodes of the graph and aggregates the entries inside the $k$ cells in the partition to return a $k\times k$ matrix. This is achieved simply by the pre- and post-multiplication 
$\mathcal S(\cdot) = \cling^\top (\cdot) \, \cling$.
Applying $\mathcal S$ to the autocovariance matrix~\eqref{eq:autocovariance} yields:
\begin{equation}
\label{eq:autocov_mat}
    \mathcal S(\mathcal F (\mathbf{x}(t)) 
    = \cling^\top \left( \bm \Pi \mathbf P(t) -\bm{\pi}\bm{\pi}^\top \right) \cling := \mathbf{R}(t, \mathbf{C}),
\end{equation}
where $\mathbf{R}(t, \mathbf{C})$ is the \textit{clustered autocovariance} of the dynamics on the network.
Due to the linearity of the chosen dynamics and the selection operator, the clustered autocovariance matrix is also the autocovariance of the coarse-grained indicator vectors $\mathbf{y}(t) = \cling^\top \mathbf{x}(t)$~ 
\cite{Delvenne2013}.
Whence, with a slight abuse of notation, we conclude that in this case:
$$\mathcal S (\mathcal F(\mathbf{x}(t)) ) = \mathcal F (\mathcal S(\mathbf{x}(t)) ) = \mathbf{R}(t, \mathbf{C}).$$

\item The quality function in Markov Stability is chosen as $\mathcal Q (\cdot) = \text{trace}(\cdot)$, i.e., the trace of the matrix. 
When applied to the clustered autocovariance matrix $\mathbf{R}(t, \mathbf{C})$, it sums over 
the $k$ autocovariances of each group for a given time $t$.
Hence, the quality of the partition is the sum of the contributions from the individual groups.
\\

In summary, Markov Stability can be written suggestively as the following function composition:
\begin{equation}
 r(t, \mathbf{C}) = \mathcal Q \left( \mathcal S \left( \mathcal F(\mathbf{x}(t)) \right)  \right )  = \text{trace } \cling^\top \left(\bm \Pi \mathbf P(t) -\bm{\pi}\bm{\pi}^\top \right) \cling.
\end{equation}
It is easy to verify that this expression is equivalent to Eq.~\eqref{eq:stability_prob}.  To find `good' partitions, our objective will be to maximise $r(t, \mathbf{C})$ over the space of partitions $\{\mathcal P \}$ for each value of $t$.
This is usually a hard combinatorial optimisation, which is approached computationally through different heuristics. 

\end{enumerate}

\subsection{Extending the framework by using other measures}

Thinking of community detection as the composition of operators immediately opens up a 
range of extensions by varying the individual ingredients.
We discuss now some of these possibilities, focussing mainly on $\mathcal{F}$ and $\mathcal{Q}$, the inference and quality operators---the choice of dynamical process $\mathbf{x}(t)$ has been discussed at length in Refs.~\cite{Delvenne2010, Lambiotte2009, Lambiotte2014}. Considering these extensions allows us to establish further connections with other community detection methods and heuristics proposed in the literature.

\subsubsection{Varying the dynamical process $\mathbf{x}(t)$}

From our discussion above, it is clear that different dynamical processes will lead to different community detection and coarse-graining algorithms. We do not dwell on this topic here, as extended studies~\cite{Lambiotte2009, Delvenne2013, Lambiotte2014} have discussed a variety of measures based on different dynamics (discrete and continuous time), and their connection with a variety of methods and heuristics proposed in the literature.
Notably, our approach leads naturally to dynamical interpretations of different variants of modularity based on distinct statistical null models which can be understood both as stationary points of the dynamics and as centrality measures of the network.
Similarly, generalised modularity measures, such as Potts models, can be understood as linearisations / linear interpolations of Markov Stability~\cite{Lambiotte2014,Delvenne2013}. 
Many further extensions are possible and, in particular translating these ideas to non-Markovian dynamics appears to be a fruitful avenue for future research.

\subsubsection{Varying the inference operator $\mathcal F$}  
A well known issue with the covariance is the fact that its absolute value is not easy to interpret, as it depends on the magnitude of the random variables involved (e.g., a simple rescaling of the variables can increase or decrease the covariance).
A standard way to discount the magnitude of the random variables is to normalise them leading to \textit{correlation measures}.
A classic example is the Pearson correlation, in which the random variables are rescaled to have unit variance.

Accordingly, we can change the inference operator to $\mathcal F = \text{corr}(\cdot)$, where the cross-correlation operator between to random vectors $\mathbf{x_1}, \mathbf{x_2}$ is defined as:
\begin{equation}
 \text{corr}(\mathbf{x_1},\mathbf{x_2}) = \mathbf S_{x_1}^{-1/2}
 \left(\mathbb E(\mathbf{x_1x_2}^\top) -\mathbb E(\mathbf{x_1})\mathbb E(\mathbf{x_2}^\top)\right) \mathbf S_{x_2}^{-1/2}.
\end{equation}
Here $\mathbf S_{x_1}$ and $\mathbf S_{x_2}$ are diagonal matrices containing the variances of the components of the vectors $\mathbf{x_1}$ and $\mathbf{x_2}$, respectively, and $\mathbb E$ denotes the expectation operator.
Substituting the inference operator in this way, leads to a correlation-based equivalent of Markov Stability:
\begin{align}
\begin{split}
r_\text{corr}(t,\cling) =& \text{trace }\cling^\top \text{corr}\left( \mathbf{x_1}(\tau),\mathbf{x_1}(\tau\!+\!t)\right) \cling \\
=& \text{trace }\cling^\top \mathbf S^{-1/2}\left(\bm \Pi \mathbf P(t) - \bm{\pi} \bm{\pi}^\top\right)\mathbf S^{-1/2} \cling,
\end{split}
\end{align}
where (like in Markov Stability) we assume that the random diffusion process is at stationarity, and thus we need to normalise by  
$\mathbf S= \bm \Pi(\mathbf I- \bm\Pi)$, the diagonal matrix with variances of Bernoulli random variables of mean $\pi_i$ on the diagonal.

Interestingly, using correlation instead of covariance was considered by Shen et al.~\cite{Shen2010} in the context of modularity community detection for directed graphs. 
Their simplified formula for undirected networks is:
\begin{equation}
 \text{Corr. Modularity} = \sum_{\mathcal{C}} \sum_{i,j \in \mathcal{C}} \dfrac{A_{ij} - \dfrac{d_i d_j}{2w}}{ \sqrt{d_i \left(1- \dfrac{d_i}{2w} \right)}\sqrt{ d_j \left(1- \dfrac{d_j}{2w}\right)}  }.
\end{equation} 
It is easy to see that this measure (proposed in~\citet{Shen2010} from a combinatorial viewpoint) corresponds to the Markov Stability with dynamics governed by a \emph{discrete-time} random walk with transition matrix 
$\mathbf{P}(t) = (\mathbf{D}^{-1}\mathbf{A})^t $ evaluated at time $t=1$, and with the Pearson correlation as the inference operator $\mathcal{F}$:
\begin{equation*}
r_\text{corr}(1,\cling)= \text{trace }\cling^\top \mathbf S^{-1/2}\left[\bm \Pi \mathbf D^{-1} \mathbf A  - \bm{\pi} \bm{\pi}^\top\right]\mathbf S^{-1/2} \cling =  \text{Corr. Modularity}.
\end{equation*}
Indeed, in the original formulation of Markov Stability~\cite{Delvenne2010}, it was already noted that a version of Markov Stability based on correlations is related to normalised cut for graph partitioning.

\paragraph{\textbf{Other extensions:}}
Although we have focussed here on the Pearson correlation, many other metrics are possible. 
A particular area of interest is the investigation of inference operators $\mathcal F$ based on mutual information.
It is also worth remarking that a number of variants arise from changing the order of application of the selection and inference operators, which will not commute in general $\mathcal{S}(\mathcal{F}(\cdot)) \neq \mathcal{F}(\mathcal{S}(\cdot))$, leading to different notions of clusterings. These directions will be the focus of future research.

\subsubsection{Varying the selection and summary operator $\mathcal S$}
The choice of operator $\mathcal{S}$ opens up a broad range of directions, which are beyond the scope of this chapter. We mention only one direction. Our focus here has been on \textit{hard} graph partitions, i.e., forcing a hard split of the nodes into non-overlapping groups. We could however allow for overlapping (possibly probabilistic) membership. 
In terms of $\mathcal S$, this would amount to relaxing the assumptions of the indicator matrix $\cling$---instead of requiring that $C_{ij} =1$ if node $i$ belongs to community $j$ and zero otherwise, we may merely require that $\sum_j C_{ij}= 1$. In that case, the node $i$ may belong to multiple groups with a certain probability.  This relaxation is equivalent to requiring that the matrix $\cling$ is row-stochastic, instead of binary, thus opening the possibility of using different optimisation techniques. 

\subsubsection{Varying the quality function $\mathcal Q$}
We now consider in some detail the choice of the quality function $\mathcal Q$,  for which an array of possibilities is also available.
Our algorithmic framework with the standard choices for Markov Stability for the operators are $\mathcal F = \text{cov}(\cdot)$, $\mathcal S = \cling^\top(\cdot) \cling$ on a diffusion dynamics $\mathbf{x}(t)$ on the network 
leading to the clustered autocovariance matrix $\mathbf{R}(t, \mathbf{C})$ given in~\eqref{eq:autocov_mat}.
To this matrix, we then apply the quality function $\mathcal Q (\cdot) = \text{trace}(\cdot)$, which sums over all the community autocovariances, a measure akin to taking an average of the individual autocovariances.

There are of course other options for the quality function. For instance, one could take into account the off-diagonal terms of the matrix $\mathbf{R}(t, \mathbf{C})$ and define a metric that searches for maximally diagonal matrices. Such an alternative operator quality function would be especially relevant in conjunction with a different, nonlinear  inference operator $\mathcal{F}$ (e.g., mutual information).
Another option is to define the quality of the partition in terms not of the average autocovariance over the groups, but of the `weakest' group autocovariance of the partition, i.e., we would favour partitions where no `bad' groups exist. 
This could be achieved by using the quality function $$\mathcal Q = \min \{\text{diag}( \cdot)\}.$$  
We consider this quality function in more detail as it provides an interesting connection with another measure for community detection proposed by Piccardi~\cite{Piccardi2011}, as we now show.

Let us consider an unbiased random walk discrete-time dynamics $\mathbf{x}(t)$ with transition matrix 
$\mathbf{P}(t) = (\mathbf D^{-1} \mathbf A)^t$ and stationary distribution $\bm{\pi} = \mathbf{d}/2w$ and define the inference operator to be $\mathcal F = \text{cov}(\cdot)$. To make the connection clearer and to avoid a bias towards partitions that contain few communities of large sizewe need to set our selection operator to perform a normalised block-averaging 
$$ \mathcal{S}(\cdot)= (\cling^\top \bm \Pi \cling)^{-1} \cling^\top (\cdot) \cling, $$
where $(\cling^\top\bm\Pi \cling) = \text{diag} ([\pi_{\mathcal{C}_1}, \ldots, \pi_{\mathcal{C}_k}])$ and 
$\pi_{\mathcal{C}_i} =  \sum_{\mathcal{C}_i} d_i/2w$ in this case. 
Thus, the $\mathcal{S}$ operator normalises each group according to the starting probability of the diffusion process within each community.
Combining these operators gives us a clustered normalised autocovariance matrix, to which we then apply the quality function 
$\mathcal Q = \min \{\text{diag}( \cdot)\}$ to obtain the following clustering measure:
\begin{equation}
\label{eq:rmin_def}
 r_\text{min}(t,\cling) = \min_i \left( (\cling^\top \bm \Pi \cling)^{-1} \cling^\top \left( \bm \Pi \mathbf P(t) -\bm{\pi}\bm{\pi}^\top \right) \cling \right)_{ii}.
\end{equation}

Piccardi \cite{Piccardi2011} proposed a technique for community detection based on the idea that the outflow from a community under a random walk should be small. To that end, he defined a lumped Markov chain with transition matrix:
\begin{equation}
\label{eq:lumped_Piccardi}
 \mathbf U = \left(\text{diag}(\bm{\pi} \cling)\right)^{-1}\cling^\top \bm \Pi \mathbf D^{-1} \mathbf A \cling,
\end{equation}
and stationary distribution $\bm{\pi}_\ell = \bm{\pi} \cling $, i.e., the 
appropriately summed version of the stationary distribution of the original Markov process.
The best split into communities (denoted an $\alpha$-partition) is found through the following optimisation
\begin{align}
\max_\cling k  \quad \text{subject to} \quad U_{ii} \geq \alpha, \quad i =1,2,\ldots,k
\end{align}
where $\alpha \in [0,1]$ is a parameter set \textit{a priori} and defining the minimal amount of flow retained in each community.

It can be shown that a partition $\mathbf{C}$ with $r_\text{min}(1,\cling) \ge \beta_1$ is also an $\alpha$-partition with a guaranteed value of $\alpha$. This follows directly from the definitions~\eqref{eq:rmin_def}~and~\eqref{eq:lumped_Piccardi}:
\begin{align}
r_\text{min}(1,\cling) \ge \beta_1  \implies 
 U_{ii} \geq \beta_1 + [(\cling^\top\bm \Pi \cling)^{-1} \cling^\top\bm \pi^\top \bm \pi \cling]_{ii}
= \beta_1 + \pi_{\ell,i} := \gamma_i.
\end{align}
Therefore a partition with $r_\text{min}(1,\cling) \ge \beta_1$ is also an $\alpha$-partition with $\alpha = \max_i \gamma_i$,
and optimising $r_\text{min}(t,\cling)$ in the space of all partitions for any $t$ encompasses $\alpha$-partitions as a special case (for $t=1$). 
For a more detailed discussion of these connections, see~\cite{Schaub2014}.

\section{Discussion} 
This chapter has focussed on the relationship between structure and dynamics in complex networks, concentrating on how the notions of time scale separation, external equitable partitions, and structural balance are related to coarse-graining and community detection in networks. 
We have exemplified these concepts through consensus dynamics and random walks, both of which have been studied extensively linked to social network analysis.

Our discussion is underpinned by the fact that both time scale separation and EEPs are intimately related to the algebraic notion of invariant subspaces. Such invariance provides us with a better understanding of the emerging dynamics on a network, and highlights structural features in the network that are of dynamical importance.
For instance, in a diffusion process time scale separation implies regions in the network where the diffusion is trapped far longer than one would expect. Hence, over a particular time scale, a diffusive signal emanating from a node inside such a group will likely reach nodes inside the same group, so that these nodes are \emph{dynamically} almost decoupled from the rest of the network.
In another example, the presence of certain graph symmetries in the network implies the presence of EEPs, graph partitions that are invariant under the dynamics. As a consequence, the averages of the consensus dynamics over the cells of the EEP can be described by the reduced dynamics of the lower-dimensional quotient graph.

Alternatively, instead of looking only at the impact of the structure on the dynamics, we can also ask the opposite question: can we use information gathered from dynamics taking place on the network to reveal important structures in the graph? 
We have proposed an algorithmic perspective to exploit this dynamics-driven route to the analysis of network structure and have shown that, interestingly, by adopting such a viewpoint we recover many methods of community detection proposed in the literature from seemingly different perspectives. 
This dynamical interpretation for such diverse methods enables us to place them with a unifying framework, providing additional insights that might not have been apparent from their initial definition.

\subsubsection*{Open directions}
There is a large literature on complex networks and dynamical processes acting on top of those, and hence there are many topics of interest related to the interplay between structure and dynamics that we have not discussed here.

A fruitful area to extend these ideas is to go beyond (first order) Markovian and diffusion-based dynamics. 
Indeed, many dynamics relevant to real world phenomena, such as epidemic spreading, are not of a diffusive type, yet
being able to obtain dynamically important modules in such systems is crucial, e.g., to design effective vaccination strategies.
Considering higher-order or non-Markovian models, allowing for the presence of memory, would for instance allow us to study bursty dynamics, or other path-dependent dynamical processes~\cite{Delvenne2015}.

Furthermore, it will be important to connect the concept of dynamical modules considered here in connection with model order reduction tools formally studied in control theory. Such a link will allow us to quantify how far the modules describe the original dynamics in a precise manner. 

Ultimately, a key aim is to describe not only the dynamics on the network in a modular fashion, but also to take into account changes of the network itself, and interactions between different types of dynamics acting on the network.
Indeed, a key area for future work (barely considered here) is that of systems where the topology of the network changes, as in social contact networks, and the inclusion of multiple kinds of interactions within one system, formalised mathematically through multiplex networks~\cite{Boccaletti2014,Kivela2014}.

\section*{Acknowledgements}
\noindent 
MTS gratefully acknowledges funding from the European Union's Horizon 2020 Research and Innovation Programme under the Marie Sklodowska-Curie grant agreement No 702410.
Most of this work was completed while MTS was at Universit\'e catholique de Louvain and University of Namur.
JCD and RL acknowledge support from: FRS-FNRS;  the Flagship European Research Area Network (FLAG-ERA) Joint Transnational Call ``FuturICT 2.0''; and the ARC (Action de Recherche Concert\'ee) on Mining and Optimisation of Big Data Models funded by the Wallonia- Brussels Federation.
MB acknowledges funding from the EPSRC, through the award EP/N014529/1 funding the EPSRC Centre for Mathematics of Precision Healthcare. The funders had no role in the design of this study. The results presented here reflect solely the authors' views.

\bibliography{references}

\end{document}